 \documentstyle[prl,twocolumn,aps]{revtex}
\begin{document}
\draft
\preprint{\bf{Applied Physics Report 93-57; cond-mat/9402031}}
\title{
Spectroscopy of the Potential Profile in a Ballistic Quantum Constriction
}
\author{
I. E. Aronov\cite{IA}, M. Jonson, and A. M. Zagoskin
}
\address{
Department of Applied Physics, Chalmers University of Technology and
University of G\"{o}teborg, S-412 96 G\"{o}teborg, Sweden
}
\maketitle
\begin{abstract}
We present a theory for the nonlinear current-voltage characteristics of a
ballistic quantum constriction. Nonlinear features
first develop because of above-barrier reflection from the potential profile,
created by impurities in the vicinity of the constriction.
The nonlinearity appears on a small voltage scale and
makes it possible to determine distances between impurities as well as the
magnitude of the impurity potentials.
\end{abstract}
\pacs{72.10.Bg, 05.60.tw, 73.40.Kp  }


In recent experiments on microconstrictions, electrostatically created by a
split-gate technique in the  2D electron gas of a GaAs-AlGaAs heterostructure,
several groups \cite{Pepper,Danmark} have reported small-scale
nonlinear features of the system conductance when measured as function of the
driving voltage, $V$. These anomalies, which occurred on the first fundamental
$2e^{2}/h$- conductance plateau, still lack a satisfactory explanation. In this
Letter we show that the small-scale nonlinearity in the conductance is
decisively
determined by the shape of the effective potential barrier
inside the quantum ballistic constriction (QBC). The latter is defined by the
geometry of the microconstriction and by impurities in its vicinity. The
effective potential profile leads to an  `above-barrier' reflection effect
\cite{LLIV}, which gives rise to the nonlinearity in the QBC under
consideration.
In short, the characteristic scale of the nonlinear features is determined by
the characteristic inter-impurity distances.

Our starting point is the  well known adiabatic model \cite{GLKhSh1988} of
a QBC shown in Fig.~\ref{fig1}. Numerical calculations \cite{Davies,Sergey}
confirm
that the effect of charged impurities in the vicinity of the microconstriction
is to smoothly modulate the electrostatic potential inside the constriction,
$V_{imp}(x,y)$. This modulation can be thought of as a simultaneous  modulation
of the microconstriction geometry {\em and}  a smooth variation of the
effective one-dimensional potential for longitudinal transport along the  QBC
axis. Such a picture is valid as long as the condition $|\partial
V_{imp}/\partial y|\  d(0) \ll \Delta E_{\perp}$ is satisfied. Here $\Delta
E_{\perp} \simeq \pi^{2}\hbar^{2}/2m^*d(0)^{2}$ is a characteristic intermode
level
spacing in the adiabatic quantum constriction.

Assuming the constriction to be long and smooth enough, the
criterion $d'(x) \simeq d(0)/l \ll 1$ is fullfilled ($l$ is the length of the
contact), and we are in the
adiabatic transport regime. Under these conditions
--- to zero order in corrections to adiabatic transport ---  the
Schr\"odinger equation for the longitudinal part of the wave function in the
{\em n-}th  transverse mode is
\begin{equation}
 -\frac{\hbar^{2}}{2m^*}\psi_{n}''(x) + W_{n}(x)\psi_{n}(x) = E\psi_{n}(x).
\end{equation}
Here $W_{n}(x) = V_{imp}(x,0) + E_{n\perp}(x)$ corresponds to an effective
gate-generated potential profile modulated
by the impurity potential both directly ($V_{imp}(x,0)$) and  through the
modulation of the constriction width, which depends on the transverse component
of the impurity potential.

We consider an electron in the $n-$th propagating mode, and assume that the
electron energy is everywhere much higher than the effective potential, $E
\gg W_{n}(x)$.   In this case above-barrier scattering will lead to a
reflection coefficient of the form \cite{LLIV}
\begin{equation}
R_{n}(E) =
\left| m^*/(\hbar^{2}k(E)) \int_{-\infty}^{\infty}W_{n}(x) e^{2ik(E)x} d\!x
\right|^{2}, \label{R}
\end{equation}
where $k(E) = \sqrt{2m^*E}/\hbar$.

The current through the contact is given by \cite{Imry}
\begin{eqnarray}
I(V) = (2e/ h) \int dE\left[n_{F}(E-\mu_{1}) \right.\nonumber \\
\left.-n_{F}(E-\mu_{2})\right]
\sum_{n=1}^{N}\left(1-R_{n}(E)\right). \label{I(V)}
\end{eqnarray}
Here $N$ is a number of  propagating modes, $\mu_{1}=E_{F}-\beta eV$ and
$\mu_{2}=E_{F}+(1-\beta)eV$ are the chemical potentials of the left- and right
electron reservoirs, and $0 \leq \beta \leq 1$ is factor that describes the
asymmetry of the voltage drop along the constriction \cite{PepperJPhys}.

The most spectacular situation occurs at zero temperature, when a simple
expression for the differential conductance of the system follows
from  Eq.~(\ref{I(V)}). One finds that
\begin{eqnarray}
G(V) = N(2e^{2}/h) - \delta G_{N}(V),
\end{eqnarray}
 where
\begin{eqnarray}
\delta G_{N}(V) = (2e^{2}/h)\sum_{n=1}^{N}\left[ \beta R_{n}(E_{F}-\beta eV)
\right.\nonumber \\
\left. +  (1-\beta)R_{n}(E_{F}+(1-\beta)eV) \right]. \label{dG}
\end{eqnarray}

Clearly,  the reflection coefficient (\ref{R}) is determined by two
factors; first by the contact geometry, and second by the impurity potential.
The geometric contribution to $W_{n}(x)$, i.e. $E_{n\perp}(x)$, is a sum of
a smooth function and a small-scale oscillatory term. The former is due to the
gate potential, while the latter is induced by the transverse part of the
impurity potential. It is well known that the smooth part of $E_{n\perp}(x)$
contributes an exponentially small term to $R_{n}$ \cite{GLKhSh1988}.
Nevertheless, as the number of propagating modes increases, the total amplitude
due to this contribution grows as   $N^{5}$ (as  $E_{n\perp} \propto
n^{2}$). Therefore, if only the number of open modes becomes large enough, the
relative amplitude of the oscillatory component of the reflection coefficient
is
weakened. The best conditions for  doing `potential profile spectroscopy' are
then at hand when the number of propagating modes is as small as possible.

{}From now on we take $N=1$.  We can thus neglect the contribution to $R_{n}$
from the adiabatic part of the transverse energy that enter $W_{n}(x)$.
Evidently, the  other contributions to this function,  coming from  the
irregularities of the constriction geometry and from the potential profile
inside the constriction, can not be separated. This
corresponds to the physical situation, since both these contributions have the
same origin: the random potential of the adjacent impurities. We denote this
part of the potential $W_{n}(x)$ --- which is independent of mode index $n$ ---
by $\tilde{V}_{imp}(x).$

Note that by considering only a single propagating mode we automatically
exclude  opening/closing of conducting modes
\cite{Pepper,Alex,First} as well as coherent mode mixing \cite{ZS}
as possible mechanisms for the nonlinear features discussed.

The effective potential $\tilde{V}_{imp}(x)$ is written in the form
\begin{equation}
\tilde{V}_{imp}(x) = \sum_{\alpha} U_{\alpha}(|x-x_{\alpha}|),
\end{equation}
where $ U_{\alpha}$ denotes the contribution to the self-consistent effective
potential from the $\alpha$-th impurity, situated at the point $x_{\alpha}$.
The reflection coefficient then acquires the simple form
\begin{eqnarray}
R(E) &=& \left(m^{*}/\hbar^{2}k(E)\right)^{2} \left|
\sum_{\alpha}e^{2ik(E)x_{\alpha}}u_{\alpha}(E)\right|^{2}, \label{spektr} \\
u_{\alpha}(E) &=& \int_{-\infty}^{\infty}dx e^{2ik(E)x} U_{\alpha}(|x|)
\nonumber
\end{eqnarray}
where $u_{\alpha}(E)$ is the Fourier transform of the effective impurity
potential.

It is evident from (\ref{spektr}) that the reflection coeffitient is sensitive
to the interimpurity distances. It follows from the uncertainty condition for
the direct and inverse Fourier transform that the accuracy $\Delta x$ with
which
these distances can  be determined from (\ref{spektr}) is limited by $\Delta x
\Delta k \geq 1/4\pi$. In our case this reduces to \begin{equation} \Delta x
>\simeq \lambda_{F} E_{F}/(4\pi^{2}|eV|). \label{dx} \end{equation} The small
numerical factor $1/4\pi^{2}$ allows for relatively small  driving voltages
to define the interimpurity distances with accuracy of order $\lambda_{F}$.

It is easy to estimate from Eq.(\ref{spektr}) the order of magnitude of the
relative conductance variation due to above-barrier scattering at zero
temperature:

\begin{eqnarray}
\frac{\delta G}{G} \simeq  \left(\pi \frac{l_{imp}}{\lambda_{F}}
\frac{ U_{imp}}{E_{F}}\right)^{2}.
\label{est}
\end{eqnarray}

Because of the smoothness of the effective impurity potential the quantity
$l_{imp} U_{imp}$ is the maximal value of its Fourier transform, $l_{imp}$
being
its effective length, and $U_{imp}$ its magnitude. The characteristic scales
for these quantitites are given by different authors in a rather wide interval.
Following \cite{Sergey}, we use $U_{imp}/E_{F} \simeq 0.3$;
$l_{imp}/\lambda_{F}$ can be estimated  to be in the range 0.1 to 0.01.
That is, the magnitude of  the nonlinear contribution to the conductance at
zero
temperature is $1-10^{-2}$\%.

At finite temperatures the conductance oscillations   will be  washed out.
Thus in order to observe the effect described it is necessary that $k_BT \ll
e\Delta V$, where $\Delta V$ is the period of oscillation with bias
voltage. A realistic estimate is that the effect can be observed if  $T \leq
100$mK.

Numerical calculations of the conductance vs. bias voltage  are presented
in Figs.~\ref{fig.3}-~\ref{fig.5}. We  use  Coulomb impurity potentials,
\begin{equation}
U_{\alpha}(|x-x_{\alpha}|) =
Z_{\alpha}e^{2}/\kappa\sqrt{(x-x_{\alpha})^{2} + h_{\alpha}^{2}}.
\label{Coulomb}
\end{equation}
 The second term under the square root determines the lateral and vertical
displacement of the $\alpha$-th impurity from the axis of the constriction. The
dielectric constant for GaAs is $\kappa \approx 13$. An
asymmetric voltage drop ($\beta \neq 0.5$) leads to an asymmetric response
$\delta G(V)$.  Evidently the larger the inter-impurity distance, the more
distinct is the  picture of oscillations, and the larger is the accuracy with
which we can determine the potential profile (see Eq.(\ref{dx})).

The average value of the nonlinear term in the conductance is described by the
averaged value of the function $R(E)\cdot E$ (see Eq.~(\ref{R})). Making use of
the Parseval inequality, we obtain the result
\begin{equation}
\left< R(E)E \right>   = \frac{1}{2\pi}\int_{-\infty}^{\infty}dk R(E)E =
\frac{m^*}{4\hbar^2}\int_{-\infty}^{\infty} dx \left|\sum_{\alpha}
U_{\alpha}(x)\right|^{2}.
\end{equation}
In this way we find the average magnitude of the impurity potential along the
axis of the constriction.

Since the Fourier transform of $R(k(E))E$ directly gives the set of
interimpurity distances, $\Delta x_{\alpha\alpha'}$, we can apply a filtration
procedure on
the corresponding frequencies, that is  find
\begin{equation}
\left<R E\right>_{\Delta x_{\alpha\alpha'}} =
\frac{1}{2\pi}\int_{-\infty}^{\infty}dk R(k) E(k) \exp(-2ik \Delta
x_{\alpha\alpha'}).
\label{filter}
\end{equation}
Hence we can analyze the lineshape of every peak in the trace of $\delta G(V)$,
initiated by each distinguishable impurity. In the general case, such an
analysis allows us to determine the quantities
$\left|h_{\alpha}-h_{\alpha'}\right|/\left|h_{\alpha}+h_{\alpha'}\right|$, thus
providing information about the distances between the impurities and the
constriction plane.

Using the model potentials (\ref{Coulomb}) we find for the line shape an
expression  $\left< R E \right>_{\Delta x_{\alpha\alpha'}} = (m^*/\hbar^2)
(e^{4}Z_{\alpha}Z_{\alpha'}/\kappa^{2}\left|h_{\alpha}+h_{\alpha'}\right|)\cdot
{\bf
 K}\left(\left|h_{\alpha}-h_{\alpha'}\right|/
\left|h_{\alpha}+h_{\alpha'}\right|\right).$
The function {\bf K}$(\xi)$ is the complete
 elliptic integral of the first
kind.

The nonlinear contribution $\delta G(V)$ to the conductance allows us to
determine the spectral  characteristics of the potential profile created by
impurities
in the constriction. For real impurity potentials
\cite{Davies,Sergey} $U_{\alpha}(k)$ are smooth functions of the variable $k$
and therefore of the driving voltage $V$. Because of this fact we can apply the
Fourier transformation technique to Eq.(\ref{dG}) and  from the positions of
the spectral peaks determine the distances $\Delta x_{\alpha\alpha'} =
|x_{\alpha}-x_{\alpha'}|$ between the impurities. Evidently, these distances
can
be  measured only if they are larger than $\lambda_{F}$. If on the contrary $
\Delta x_{\alpha\alpha'} \leq \lambda_{F}$ we can regard the corresponding
impurity cluster as a single scatterer. Anyhow, it  follows from (\ref{dG})
that
they can be resolved only at large enough voltages $V$, i.e.,  in the regime
where the nonlinear effects are developed. A Fourier analysis of
$G(V)$ as well permits us to find the quantities \cite{BOOK}
$\sum_{\alpha}|U_{\alpha}|^{2}$ and $\sum_{\alpha\neq\alpha'}
U_{\alpha}U_{\alpha'}^{*}\cos({\tt Arg}(U_{\alpha}U_{\alpha'}^{*}))$, which
characterize the magnitudes of the impurity potentials.

In Figs.\ref{fig.3}-\ref{fig.5} different types of
conductance-voltage curves caused by above-barrier reflection in the
constriction are shown. We plot there the nonlinear contribution $\delta G$
to the differential conductance(in arbitrary units) vs. the reduced driving
voltage
$v = V/E_F$. The scale of  $v \in [-0.5;0.5]$ is consistent with
measured values of $E_F \sim 10$meV and distance between the first and second
subbands in a QBC, $\Delta E_\perp \simeq 5$meV [1]. The potential drop is
assumed to be completely asymmetric $(\beta = 1)$.

 In Fig.\ref{fig.3} the next simplest case is shown,
with a single impurity. Evidently the nonlinear correction is a monotonic
function of voltage. Nonmonotonitic behavior appears when two impurities
are present (Fig.\ref{fig.4}). Fig.\ref{fig.5} demonstrates
the case of three impurities. It is easy to see that the structure of the
curves becomes
richer as the number of impurities and/or average distance between them grows.

In conclusion, we have demonstrated that above-barrier scattering of
electrons by the potential profile in quantum ballistic constrictions leads to
small-scale nonlinearities in the current-voltage characteristics. The effect
is
best observed  when only one conducting channel is open, and at temperatures
less
than $\simeq 100$ mK. It can be used to determine the spatial arrangement of
the
impurities in the constriction.


This work was supported by the Swedish Academy of Sciences (KVA), the Natural
Science Research Council (NFR), and by NUTEK. We are grateful to P. E.
Lindelof, L. Kuzmin, M. Persson, and R. I. Shekhter for productive discussions.
One of us (IA) acknowledges the hospitality of the Department of Applied
Physics at CTH/GU.


%
%

\begin{figure}
 \caption{Quantum ballistic constriction. (a) Clean adiabatic channel; $d(x)$
is a smooth function of longitudinal coordinate. (b) Effective potential
profile
inside the channel in presence of impurities. }
 \label{fig1}
 \end{figure}

\begin{figure}
 \caption{  Nonlinear contribution to the conductance of the constriction with
a single impurity.}
 \label{fig.3}
 \end{figure}

\begin{figure}
 \caption{  Nonlinear contribution to the conductance of the constriction with
two impurities. The oscillations arise due to Fabry - Perot type interference
of electron wave between them (the distance between the impurities is
11$\lambda_F$).}
 \label{fig.4}
 \end{figure}

\begin{figure}
 \caption{   Nonlinear contribution to the conductance calculated for 3
impurities in the constriction. The average distance
between the impurities equals to $23\lambda_{F}$ .}
 \label{fig.5}
 \end{figure}

\end{document}